# An algorithm to increase the residues of wrapped-phase in spatial domain

Guangliang Du[a], Minmin Wang[a], Canlin Zhou [a*],Shuchun Si[a], Hui Li[a], Zhenkun Lei[b],Yanjie Li[c]

[a] School of Physics, Shandong University, Jinan 250100, China

[b] Department of Engineering Mechanics, Dalian University of Technology, Dalian 116024, China

[c] School of Civil Engineering and Architecture, University of Jinan, Jinan, 250022, China

*Corresponding author: Tel: +8613256153609; E-mail address: canlinzhou@sdu.edu.cn

## Abstract

In phase unwrapping, the locations and densities of residues are indicative of the severity of the unwrapping problem. The residues are used to detect and evade inconsistent phase areas. Gdeisat et al. proposed an algorithm to increase the number of residues in a wrapped-phase map to improve the results of phase unwrapping. But this method will take much time to make the Fourier transform, inverse Fourier transform, select and shift the spectral components, and there is no theoretical analysis on why the frequency shift can increase the number of residues. In view of the above problems, we proposed an algorithm to increase the number of residues in a wrapped-phase map, which only uses a simple multiply operation in spatial domain to realize frequency shift by taking advantage of the frequency shift property of Fourier transform. Besides that, we discuss the relationship between the number of residues and frequency shift. Finally, the experimental evaluation is conducted to prove the validity of the proposed method. Experimental results demonstrated that the proposed method can speed up more than 50%.

## 1. Introduction

  Phase-based fringe projection technique is an important method in three-dimensional (3D) shape measurement. It has been extensively investigated and widely used in many fields for its simple device and higher accuracy [1-4]. Since the phase obtained by phase demodulation method is wrapped in $(-\pi,\pi)$, phase unwrapping become one of the critical steps of 3D shape measurement [5-7]. Most phase-unwrapping algorithms can be classified into two categories: temporal phase unwrapping and spatial phase unwrapping. Temporal phase unwrapping method requires two or more wrapped phase maps [8-11], while the spatial phase unwrapping method requires only one wrapped phase map to complete unwrapping. Among them, path independent phase unwrapping algorithm is an active kind of algorithms. Goldstein et al. [12] proposed the famous branch-cut method in 1988, which has been widely used. Because this algorithm is very dependent on the placement of the branch cuts, there will be a lot of errors in the unwrapping results if the noise is too much, so many people then proposed some improved methods. Flynn et.al [13] proposed the quality guided path following method in 1996. Xu [14] proposed a region-growing phase unwrapping algorithm. Fornaro [15] et al. proposed a multi-channel phase unwrapping algorithm based on phase differences in 2005. Hu proposed a color fringe projection method [16]. Zhang proposed a stair phase coding technology [17]. Chen [18] proposed a method making use of the three primary color channels associated with digital projectors. Goldstein [19] proposed a smart temporal unwrapping that temporally unwraps the phase data such that small motion between frames is accounted for and phase data are unwrapped consistently between frames. Liu [20] proposed a phase retrieval method using a composite fringe with multi-frequency. Souza [21] proposed a two-dimensional phase unwrapping algorithm based on the theory of residues which is essential in the development of branch-cut algorithms. These methods make great progress in phase unwrapping.

Recently, Gdeisat et al. proposed a method to increase the number of residues in two-dimensional phase-wrapped images that contain discontinuities [22], which can be used to improve the performance of path independent phase unwrapping algorithm such as branch-cut algorithm. However, according to our own experience with the method, Gdeisat's method has the following disadvantages: (1) In the literature, there is no theoretical analysis on why the frequency shift can increase the number of residues. (2) This method needs Fourier transform, inverse Fourier transform, select and shift the spectral components, these procedures increase the calculation complexity and the processing time consuming as well.
Here, in order to improve the calculation efficiency as well as simplify its procedures, we present an algorithm to increase the number of residues of wrapped-phase in spatial domain, which takes advantage of the frequency shift property of two-dimensional(2D) Fourier transform [23]. This method is a good solution to the problems of Gdeisat's method. The capability of the presented method is demonstrated by both theoretical analysis and experiments.

The paper is organized as follows. Section 2 introduces the principle of the system. Section 3 presents the experimental results. Section 4 summarizes this paper.

## 2. Theory
### 2.1 Gdeisat's method

In 3-D shape measurement, surface relief, shadow and phase noise may cause local errors in the unwrapping phase, Goldstein called the local errors as residues [12]. In order to detect and calculate the residue, a $2\times 2$ window is defined in a wrapped phase diagram, and then the difference between adjacent pixels(phase gradient value) is calculated by the following formula in a fixed direction which can be either clockwise or counterclockwise. Then the phase gradient value is accumulated.

$$\Delta 1 = W(\varphi_w(i, j+1) - \varphi_w(i, j))$$
$$\Delta 2 = W(\varphi_w(i+1, j+1) - \varphi_w(i, j+1))$$
$$\Delta 3 = W(\varphi_w(i+1, j) - \varphi_w(i+1, j+1))$$
$$\Delta 4 = W(\varphi_w(i, j) - \varphi_w(i+1, j))$$
$$q = \frac{(\sum_{i=1}^{4} \Delta i)}{2\pi}$$
(1)

where $W$ is the wrapping phase operator, $\varphi_w(i, j)$ is the wrapped phase at (i,j). Mark the upper left corner of the $2\times 2$ window as the center point. When q=0, the center point is not a residue; when q>0, it is a positive residue; on the contrary, it is a negative residue. In the original wrapped phase map, move the window, we can detect and locate the residues in the whole two-dimensional phase map. The locations and densities of residues are indicative of the severity of the phase unwrapping problem and graphically indicate where problems in the unwrapping process are likely to occur. The residues can be used to place branch cuts that are used as barriers to prohibit the phase-unwrapping path from passing through them. Gdeisat[22] thinks that the residues are the indicators of poor quality regions of the phase, and increase their number is helpful to detect and avoid the inconsistent phase areas. So, they proposed a method that uses the Fourier transform to increase the number of residues in the original wrapped phase map. The details for Gdeisat's method can be found in [22],The main stages of Gdeisat's algorithm are described as follows:

(1)convert the original wrapped-phase map into the complex array $\varphi_{wc}(x, y)$.

(2) make the Fourier transform to $\varphi_{wc}(x, y)$ to obtain $\Phi(u, v)$.

(3) select and move the spectrum to obtain $\Phi(u-u_0, v-v_0)$.

(4) make the inverse Fourier transform to obtain $\varphi_{wcs}(x, y) = F^{-1}[\Phi(u-u_0, v-v_0)]$.

(5) generate the new wrapped phase map from $\varphi_{wcs}(x, y)$.

(6)calculate the new residues from the new wrapped phase map.
(7)construct quality map or branch cut or mask from the new residues.
(8)unwrap the original wrapped phase by the unwrapping algorithm.

The method in ref. [22] can improve considerably the performance of the Goldstein algorithm by using new residues as mask. Experiments will be described in the third section.

**2.2 Our method**

In the literature 22, the preprocessing method to increase the number of residues can improve the accuracy of the quality guided or branch-cut phase-unwrapping algorithm, but it also has some disadvantages, that is, (1) there is no discussion on why the frequency shift can increase the residues. (2) this method needs twice Fourier transform, and it needs to search and select the spectral components, which will occupy large amount of processing time. Based on our analysis of the ref. [22], we put forward the corresponding solutions. The basic idea of our method is that the complex procedures such as Fourier transformation, frequency selection and inverse transformation are not required, which are replaced by a simple multiply operation in spatial domain.

In this paper, we only use the four-step phase shift algorithm to obtain the wrapped phase. However, the method proposed in this paper also can be used to process wrapped phase maps that have been extracted using other algorithms such as three-step phase shift algorithm.

Assuming that the four phase-shifted fringe images is as follows,

$$I_1(x,y) = \cos[2\pi f_x x + 2\pi f_y y + \beta\varphi(x,y)]$$
$$I_2(x,y) = \cos[2\pi f_x x + 2\pi f_y y + \beta\varphi(x,y) + \frac{\pi}{2}]$$
$$I_3(x,y) = \cos[2\pi f_x x + 2\pi f_y y + \beta\varphi(x,y) + \pi] \quad (2)$$
$$I_4(x,y) = \cos[2\pi f_x x + 2\pi f_y y + \beta\varphi(x,y) + \frac{3\pi}{2}]$$

Where $f_x$ and $f_y$ are the spatial carrier frequency along the X axis and the Y axis respectively, $\beta$ is the modulation index. The wrapped phase can be extracted using the well-known Eq.(3) below by the four-step phase shift algorithm.

$$\varphi_w(x,y) = \tan^{-1}\left[\frac{I_4 - I_2}{I_3 - I_1}\right] = W(2\pi f_x x + 2\pi f_y y + \beta\phi(x,y)) \quad (3)$$

Where $\tan^{-1}$ is the four quadrant arctangent operator, and $\varphi_w(x,y)$ is the wrapped phase which is wrapped in $(-\pi, \pi)$. The following three equations convert the wrapped phase map into the complex array $\varphi_{wc}(x,y)$.

$$R(x,y) = \cos[\varphi_w(x,y)]$$
$$I(x,y) = \sin[\varphi_w(x,y)] \quad (4)$$
$$\varphi_{wc}(x,y) = R(x,y) + jI(x,y)$$

Where $j$ is equal to $\sqrt{-1}$.

In Gdeisat's method, first, make the Fourier transform to $\varphi_{wc}(x,y)$ as shown in Eq.(5).

$$\Phi_{wc}(u,v) = \xi[\varphi_{wc}(x,y)] \quad (5)$$

where $\xi[.]$ is the 2D Fourier transform operator, and the terms $u$ and $v$ are the vertical and horizontal frequencies respectively. The 2D Fourier transform of the wrapped phase $\Phi_{wc}(u,v)$ is shifted away from the origin using the indices $u_0$ and $v_0$ (the frequency shift should not exceed a quarter of the image size [22]), then do the inverse Fourier transform.

$$\varphi_{wcs}(x,y) = \xi^{-1}[\Phi_{wc}(u-u_0, v-v_0)] \qquad (6)$$

where $\xi[.]^{-1}$ is the inverse 2D Fourier transform operator, $u_0$, $v_0$ is moving distance. We can obtain the phase map as follows.

$$\varphi_{ws}(x,y) = \tan^{-1}\frac{I\{\varphi_{wcs}(x,y)\}}{R\{\varphi_{wcs}(x,y)\}} = W(2\pi(f_x+u_0)x + 2\pi(f_y+v_0)y + \beta\varphi(x,y)) \quad (7)$$

where $I\{.\}$ represents the imaginary part, and $R\{.\}$ represents the real part of the complex array $\varphi_{wcs}(x,y)$.

From Eq.(7), it is not difficult to understand that why the frequency shift can increase the number of residues in the wrapped phase map. Shifting the spectrum in the frequency domain equals to increasing the spatial carrier frequency of projected fringe, thus increase the number of phase wraps in the phase map. Therefore, the sensitivity of projected fringe pattern is enhanced and the number of residues become more.

By analyzing the implementation process in the ref. [22], we can see Gdeisat's method requires twice Fourier transform, once spectrum search and selection (If we select spectrum with an interactive approach in the frequency domain, it is not suitable for the automatic implementation of the algorithm. If we use the automatic method to search for peak in the frequency domain, it is necessary to design a band-pass filter window after locating the peak position, the process is relatively cumbersome.), and finally it need to move the spectrum. Therefore, in the actual operation, the process is relatively complex. To solve this problem, We simplify and speed up the implementation process by the frequency shift property of 2D Fourier transform. As shown in Ref. [23], the frequency shift property of 2D Fourier transform can be written as:

$$F(u-u_0, v-v_0) = \xi[f(x,y)e^{(j2\pi(u_ox/m+v_0y/n))}] \qquad (8)$$

where $F(u,v)$ is the Fourier transform of $f(x,y)$, $m$, $n$ are the length of $f(x,y)$ along the X axis and the Y axis respectively.

By Eq.(8), we can directly obtain $\varphi_{wcs}(x,y)$ by multiplying $e^{(j2\pi(u_ox/m+v_0y/n))}$ by the result $\varphi_{wc}(x,y)$ of Eq.(4).

$$\begin{aligned}\Phi_{wc}(u-u_0, v-v_0) &= \xi[\varphi_{wc}(x,y)e^{(j2\pi(u_ox/m+v_0y/n))}] \\ \varphi_{wcs}(x,y) &= \xi^{-1}[\Phi_{wc}(u-u_0, v-v_0)] = \varphi_{wc}(x,y)e^{(j2\pi(u_ox/m+v_0y/n))}\end{aligned} \qquad (9)$$

By Eq.(9), we can realize frequency shift in spatial domain by the frequency shift property of 2D Fourier transform. It is obvious that our method does not require Fourier transformation, spectrum selection, spectrum shift and inverse Fourier transformation. Thus, the proposed method can greatly save the computing time.

The phase $\varphi_{ws}(x,y)$ can be obtained directly by calculating the phase angle of the complex number $\varphi_{wcs}(x,y)$.

The main stages of our algorithm are summarized as follows:

(1) convert the original wrapped-phase map into the complex array $\varphi_{wc}(x,y)$.

(2) multiply $\varphi_{wc}(x,y)$ by $e^{(j2\pi(u_o x/m + v_0 y/n))}$ to obtain $\varphi_{wcs}(x,y)$.

(3) generate the new wrapped phase map from $\varphi_{wcs}(x,y)$.

(4) calculate the new residues from the new wrapped phase map.
(5) construct quality map or branch cut or mask from the new residues.
(6) unwrap the original wrapped phase by the unwrapping algorithm.

The following experiment is used to verify the proposed algorithm.

## 3. Experiments

In this section, for evaluating the real performance of our method, we test our method on a series of experiments. Below, we will describe these experiments and practical suggestions for the above procedure.

We develop a fringe projection measurement system, which consists of a DLP projector (Optoma EX762) driven by a computer and a CCD camera ( DH-SV401FM). Fig. 1 shows the schematic of fringe-projection profilometry system, where P is the projection center of the projector, C is the camera imaging center, and D is an arbitrary point on the tested object. The captured image is 768 pixels wide by 576 pixels high. The surface measurement software is programmed by MATLAB with I5-4570 CPU @ 3.20 GHz.

The tested object is a hand. Fig. 2 shows the captured image, and the wrapped phase obtained by four-step phase shift algorithm is shown in Fig. 3.

Fig. 4 shows the location of residues in the wrapped-phase map by [12], there are a total of 47 residues.

Fig.5 shows the unwrapped phase by branch-cut phase-unwrapping method [12], we can see that some of the results are incorrect.

Fig.6 is the wrapped phase obtained by Gdeisat's method and the proposed method, that is say, the frequency of 2D Fourier transform spectrum is shifted using values $u_0=50, v_0=50$ in frequency and spatial domain respectively. The results processed by both method are same. Compared with Fig.6 and Fig.3, we can see, as the spectrum moves, the carrier frequency becomes larger, so the number of phase wraps in Fig.6 are increased significantly.

Fig. 7 shows the location of residues in Fig.6 by [12], there are a total of 77 residues. The number of residues are increased.

The wrapped phase map shown in Fig.3 is unwrapped again by branch-cut phase-unwrapping method [11,22], but this time these residues shown in Fig.7 used as mask that is provided to guide the branch-cut phase-unwrapping method [12,22]. The resulting unwrapped phase map is shown in Fig.8. Comparing both unwrapped phase in Fig.8 and Fig.5 demonstrated that the increase the number of residues can improve the accuracy of branch-cut phase-unwrapping method.

In order to compare the time between Gdeisat's method and the proposed method, we conduct a comparative experiment.

Table 1 lists the comparisons of time consuming between these two methods. In the comparisons, all processed fringe patterns have the pixels sizes of 768 x 576, the computational platform is a personal laptop with Intel Core i5-4570 CPU at 3.20GHz and a 4GB RAM. We use MATLAB 2014a on the same computer to process the same fringe pattern.

| method | Time consuming |
| --- | --- |
| Gdeisat's method | 1.460s |
| The proposed method | 0.712s |

Table.1.   Comparisons of time consuming of the two methods

As can be seen from table 1, the difference of time consuming between Gdeisat's method and the proposed method is obvious, the speed of the proposed algorithm is improved by about 50%. In the proposed method, the calculation procedures such as Fourier transformation, frequency selection and inverse transformation are not required, therefore, it can save large amount of processing time.

## 4. Conclusion

In this paper, we propose an algorithm to increase the number of residues of wrapped-phase in spatial domain which is an extension of Gdeisat's method. Our method overcomes the main disadvantages that Gdeisat's method encounters. The proposed method eliminates Fourier transformation, inverse Fourier transformation and frequency selection in Gdeisat's method, and achieves the frequency shift by a simple multiply operation in spatial domain. So the proposed method can save large amount of processing time. Experimental results demonstrated that the proposed method can speed up more than 50%.

**Acknowledgment**
This work was supported by the National Natural Science Foundation of China (Grant nos. 11302082 and 11472070). The support is gratefully acknowledged.

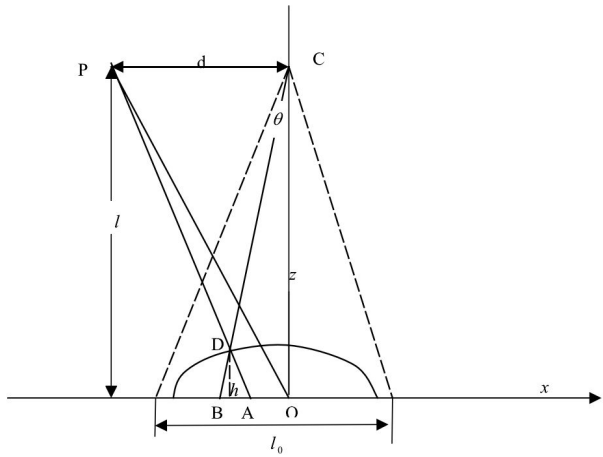

Fig.1.Optical path of phase measuring profilometry

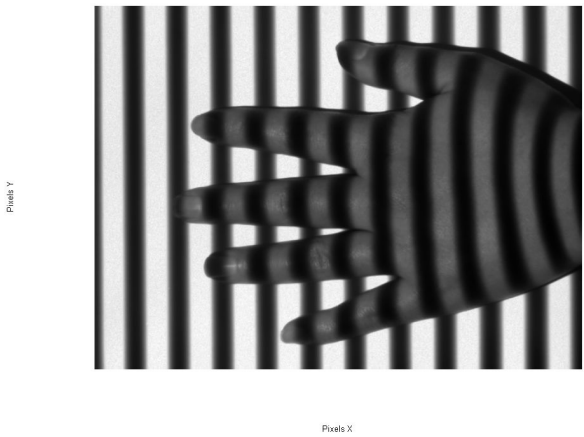

Fig. 2.the first pattern of the sinusoidal fringe

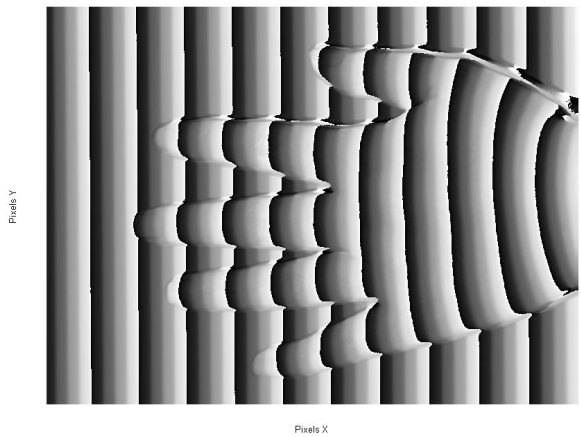

Fig. 3. the initial wrapped phase

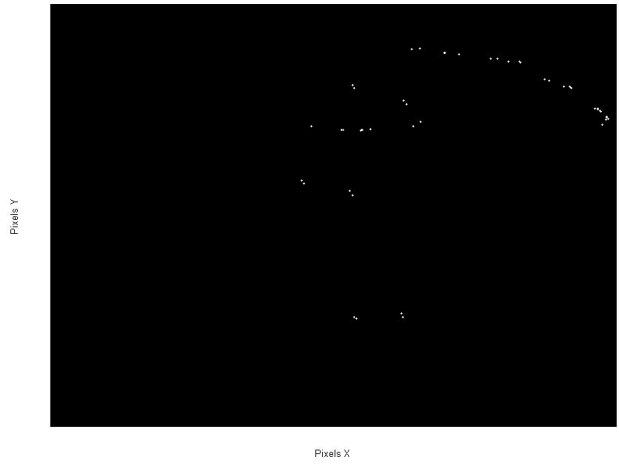

Fig. 4. the location of residues in the wrapped-phase map

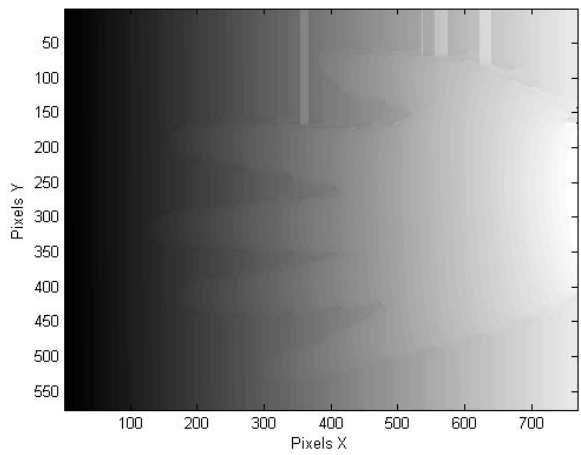

Fig.5. the unwrapped phase by branch-cut phase-unwrapping method

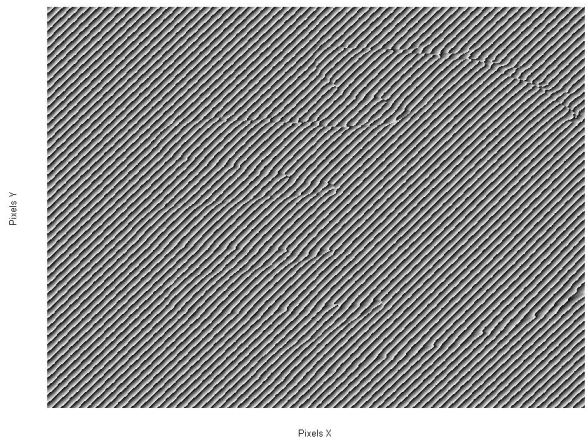

Fig.6. the new wrapped phase obtained by the proposed method($u_0$=50,$v_0$=50)

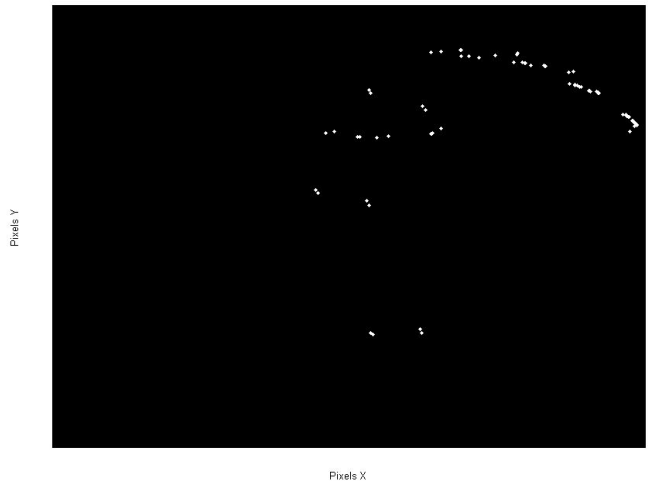

Fig.7. the location of residues in Fig.6

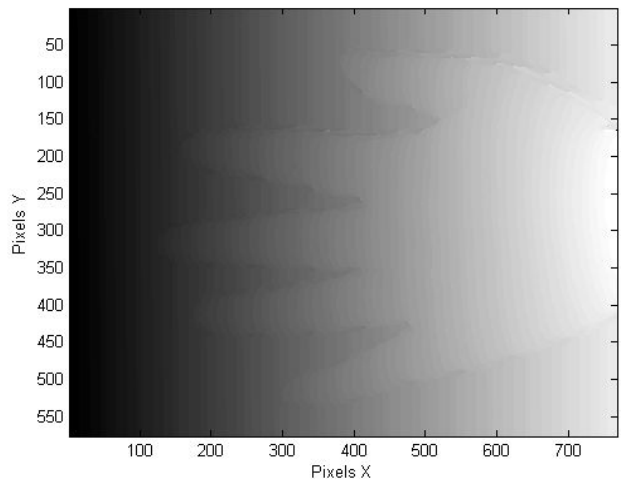

Fig.8. the unwrapped phase by branch-cut phase-unwrapping method